\documentclass[10pt,conference, a4paper]{IEEEtran} % use the `journal` option for ITherm conference style
\IEEEoverridecommandlockouts
\usepackage{cite}
\usepackage{multirow}
\usepackage{comment}
\usepackage{amsmath,amssymb,amsfonts}
\usepackage{algorithm}
\usepackage{algorithmic}
\usepackage{graphicx}
\usepackage{textcomp}
\usepackage{caption}
\usepackage{array}
\usepackage{tabularray}
\usepackage{ragged2e}
\usepackage{url}
\usepackage[labelformat=simple]{subcaption}

\usepackage[table,xcdraw]{xcolor}
\def\BibTeX{{\rm B\kern-.05em{\sc i\kern-.025em b}\kern-.08em
    T\kern-.1667em\lower.7ex\hbox{E}\kern-.125emX}}
\begin{document}
%%text separation 
%\setlength{\textfloatsep}{0.3pt}

\title{Handover Delay Minimization in Non-Terrestrial Networks: Impact of Open RAN Functional Splits}

%\begin{comment}
\author{
 \IEEEauthorblockN{Siva Satya Sri Ganesh Seeram\IEEEauthorrefmark{1}, Luca Feltrin\IEEEauthorrefmark{2}, Mustafa Ozger\IEEEauthorrefmark{1}\IEEEauthorrefmark{3}, 
Shuai Zhang\IEEEauthorrefmark{1}, Cicek Cavdar\IEEEauthorrefmark{1}}
\IEEEauthorblockA{\IEEEauthorrefmark{1} KTH Royal Institute of Technology, Sweden, Email: \{sssgse, shuai2, cavdar\}@kth.se \\
\IEEEauthorrefmark{2} Ericsson AB, Sweden, Email: luca.feltrin@ericsson.com\\
\IEEEauthorrefmark{3} Aalborg University, Denmark, Email: mozger@es.aau.dk
}
} 
%\end{comment}

\begin{comment}
\author{\IEEEauthorblockN{ Siva Satya Sri Ganesh Seeram, Shuai Zhang, Mustafa Ozger, and Cicek Cavdar} \IEEEauthorblockA{ {School of Electrical Engineering and Computer Science,} {KTH Royal Institute of Technology}, Stockholm, Sweden } \IEEEauthorblockA{E-mail: \{sssgse, shuai2, ozger, cavdar\}@kth.se} }
\end{comment}

\maketitle
%\begingroup\renewcommand\thefootnote{\textsection}
%\footnotetext{This paper does not reflect company view at the time of submission.}
%\endgroup
\begin{abstract}
This paper addresses the challenge of optimizing handover (HO) performance in non-terrestrial networks (NTNs) to enhance user equipment (UE) effective service time, defined as the active service time excluding HO delays and radio link failure (RLF) periods.  Availability is defined as the normalized effective service time which is effected by different HO scenarios: Intra-satellite HO is the HO from one beam to another within the same satellite; inter-satellite HO refers to the HO from one satellite to another where satellites can be connected to the same or different GSs. We investigate the impact of open radio access network (O-RAN) functional splits (FSs) between ground station (GS) and LEO satellites on HO delay and assess how beam configurations affect RLF rates and intra- and inter-satellite HO rates. This work focuses on three O-RAN FSs—split 7.2x (low layer 1 functions on the satellite), split 2 (layer 1 and layer 2 functions on the satellite), and gNB onboard the satellite—and two beam configurations (19-beam and 127-beam). In a realistic dynamic LEO satellite constellation where different types of HO scenarios are simulated, we maximize effective service time by tuning the time-to-trigger (TTT) and HO margin (HOM) parameters. %Using an exhaustive search algorithm, we analyze the objective function across a range of TTT and HOM values, and identify that optimal values of $\delta_t = 0$ s and  $\delta_h = 3$ dB yield the highest effective service time for each FS deployment. 
Our findings reveal that the gNB onboard the satellite achieves the highest availability, approximately $95.4$\%, while the split 7.2x exhibits the lowest availability, around $92.8$\% due to higher intra-satellite HO delays.
\end{abstract}

\begin{IEEEkeywords}
non-terrestrial network (NTN), conditional handover (CHO), open radio access network (O-RAN), low earth orbit (LEO) satellite, radio link failure (RLF)
\end{IEEEkeywords}

\section{Introduction}
\label{sec:introduction}
Deployment of low Earth orbit (LEO) satellites has become increasingly vital in providing global connectivity, particularly in remote or underserved regions where traditional terrestrial infrastructure is limited. They are characterized by their low-altitude orbits, enabling reduced latency and improved communication link quality compared to traditional geostationary (GEO) satellites. However, because LEO satellites move at high relative velocities, users experience frequent handovers (HOs) as satellites move in and out of their visibility range creating challenges in maintaining stable, high-quality connections, especially for mobile user equipment (UE).

Effectively managing these frequent HOs is essential for ensuring continuous service and minimizing latency. Prior research has proposed various strategies to address HO challenges in LEO satellite networks. For example, in \cite{LiUser20}, Xue \emph{et al.} proposed a user-centric HO mechanism to enhance the quality of service (QoS) by allowing ground users to switch to satellites with optimal downlink data, stored across multiple satellites. Another approach by \cite{LeiHandover21} uses a dynamic preference model to identify optimal satellite candidates, while \cite{JuanHandover22} evaluates HO performance in terms of unnecessary HOs (UHOs) and ping-pong (PP) events for gNodeB (gNB) based functional split (FS) deployment. In~\cite{seamlessHO}, the authors proposed a novel HO scheme in LEO-based NTN for service continuity optimization with optimal target satellite selection. In \cite{Jinxuan22}, Chen \emph{et al.} applied reinforcement learning (RL) to manage HOs for flying vehicles and ground users, considering metrics such as remaining visibility time, signal quality, and available idle channels. Authors of \cite{ZhengInter24} proposed inter-beam HO algorithms for LEO satellites using conditional HO (CHO) schemes that factor in event-based triggers and link conditions. Other studies have analyzed HO criteria in relation to performance metrics. For instance, Demir \emph{et al.} in \cite{HOperf1}, compares traditional measurement-based HOs with alternative approaches, concluding that conventional HOs can outperform alternatives when HO parameters such as offset margin and time-to-trigger (TTT) are optimized. Similarly, \cite{HOperf2} demonstrates that multi-connectivity HO can improve performance over single-connectivity HO by reducing interruptions. In~\cite{beam0}-\cite{beam3}, the multi-beam satellite modeling is illustrated for varying beam layouts and satellite configurations. However, to the best of our knowledge, no work has considered the effect of HO management on varying multi-beam satellite architectures. This paper addresses this gap by implementing and analyzing HO performance across different beam configurations, contributing new insights into the optimization of HO strategies in dynamic LEO satellite environments.

The emergence of new-generation LEO satellites with regenerative architecture enables baseband processing at the satellites with the flexibility of splitting radio access network (RAN) functions between ground station (GS) and satellites. Open RAN (O-RAN) architecture is a suitable candidate for NTNs to accommodate this flexibility in function placement with open interfaces~\cite{oran2}. In O-RAN, the RAN functions are disaggregated into open central unit (O-CU), open distributed unit (O-DU), and open radio unit (O-RU). This work considers three O-RAN FSs~\cite{oran}, namely O-RAN split 7.2x, O-RAN split 2, and gNB onboard the satellite. In the O-RAN split 7.2x, only layer 1 (L1) functions—such as beamforming, precoding, inverse fast Fourier transform (IFFT), and radio frequency (RF) processing—are placed in O-RU on the satellite, while the remaining higher layer functions are deployed on GS in O-DU and O-CU. In the O-RAN split 2, both L1 (physical layer) and layer 2 (L2) functions reside on the satellite in O-RU and O-DU, while layer 3 (L3) functions are hosted at the GS on O-CU. The gNB onboard the satellite places the entire O-RAN stack on the satellite, increasing satellite processing requirements but reducing the fronthaul bandwidth demand~\cite{eucnc}.

In our previous work~\cite{eucnc}, we analyzed the HO delay as a function of mobility scenario and FS. Different mobility scenarios arise, such as HO within beams of the same satellite (intra-satellite HO), or between two satellites connected to the same GS or different GSs (inter-satellite HO). Our snapshot analysis revealed that in intra-satellite HOs, split 7.2x incurs a higher delay (approximately $97$ ms) compared to gNB ($52$ ms). For inter-satellite HOs, split 7.2x experiences delay ranging from $97$ ms to $219$ ms, depending on the scenario, while gNB delay ranges between $173$ ms and $189$ ms. These findings indicate that the type of mobility scenario and its occurrence significantly influence overall HO delay in a dynamic LEO constellation setup.

Building on our previous work~\cite{eucnc}, this paper considers a dynamic LEO satellite constellation to assess the impact of O-RAN FSs on HO delay across varying mobility scenarios. Our objective is to jointly minimize HO delay and radio link failures (RLFs) duration by optimizing HO parameters such as TTT and HO margin (HOM). However, comprehensive analysis of the combined effects of O-RAN FSs and beam configurations on HO performance metrics—such as RLF rates, intra- and inter-satellite HO rates, and overall UE's effective service time—remains an open research issue. Effective service time is defined as the duration for which the UE is served excluding the HO delay and RLF duration.

%In this paper, we investigate the Conditional Handover (CHO) mechanism in a LEO satellite network. Ground UEs are served through a LEO satellite constellation architecture involving ground stations (GSs) and inter-satellite links (ISLs). Each satellite in this network is equipped with directional circular aperture antennas that form Earth-moving cells/beams \textit{i.e.,} non-steerable beams to serve UEs on the ground. We model and analyze HO performance, specifically focusing on inter- and intra-satellite HOs as well as CHO delays, adapting them to the dynamic nature of LEO satellite constellations.
None of the aforementioned studies has systematically evaluated the HO performance across different O-RAN FSs and beam configurations in LEO satellite networks. This paper addresses this gap by implementing a realistic LEO constellation model and employing a reference signal received power (RSRP) based CHO scheme, detailed in later sections, to study the influence of O-RAN FSs and beam configurations on the effective service time of the UE. Hence, in this paper, our main contributions are summarized as follows:

\begin{itemize}
    \item We propose a detailed HO delay model considering LEO satellite constellation and GSs under different O-RAN FSs between satellite and GSs. 
    \item We propose effective service time as a new metric considering both the HO delay and RLF duration.
    \item We formulate an optimization problem to maximize UE effective service time in NTN by jointly minimizing RLFs and total CHO delays, and we employ an exhaustive search algorithm to evaluate this objective function across a range of TTT and HOM values, identifying optimal configurations. 
    \item We analyze the impact of HO parameters like TTT and HOM on key performance metrics, including RSRP, RLF, intra- and inter-satellite HO rates, and the effects of different O-RAN FSs and beam configurations.
\end{itemize}
%Through this research, we aim to provide insights that contribute to the development of more resilient and responsive LEO satellite networks, ultimately enhancing their capacity to deliver reliable connectivity on a global scale.
The remainder of this paper is organized as follows: Section~\ref{sec:system model} presents the system model, including the path loss model, CHO delay model and key performance indicators (KPIs). Section~\ref{sec:prob form} discusses problem formulation. Section~\ref{sec:results} presents simulation results, analyzing HO performance across different beam configurations and examining the impact of O-RAN FSs on total CHO delay and effective service time. Finally, Section~\ref{sec:conclusion} concludes with a summary of findings and implications.

\section{System Model}

\label{sec:system model}
In this section, we first outline the system architecture implemented in our simulation, followed by the path loss model and its associated parameters. Next, we describe the CHO delay model and conclude with a discussion of the mobility KPIs used in our analysis.
\subsection{System Architecture}

\begin{figure}[t]
\centering
\includegraphics[width=0.99\linewidth]{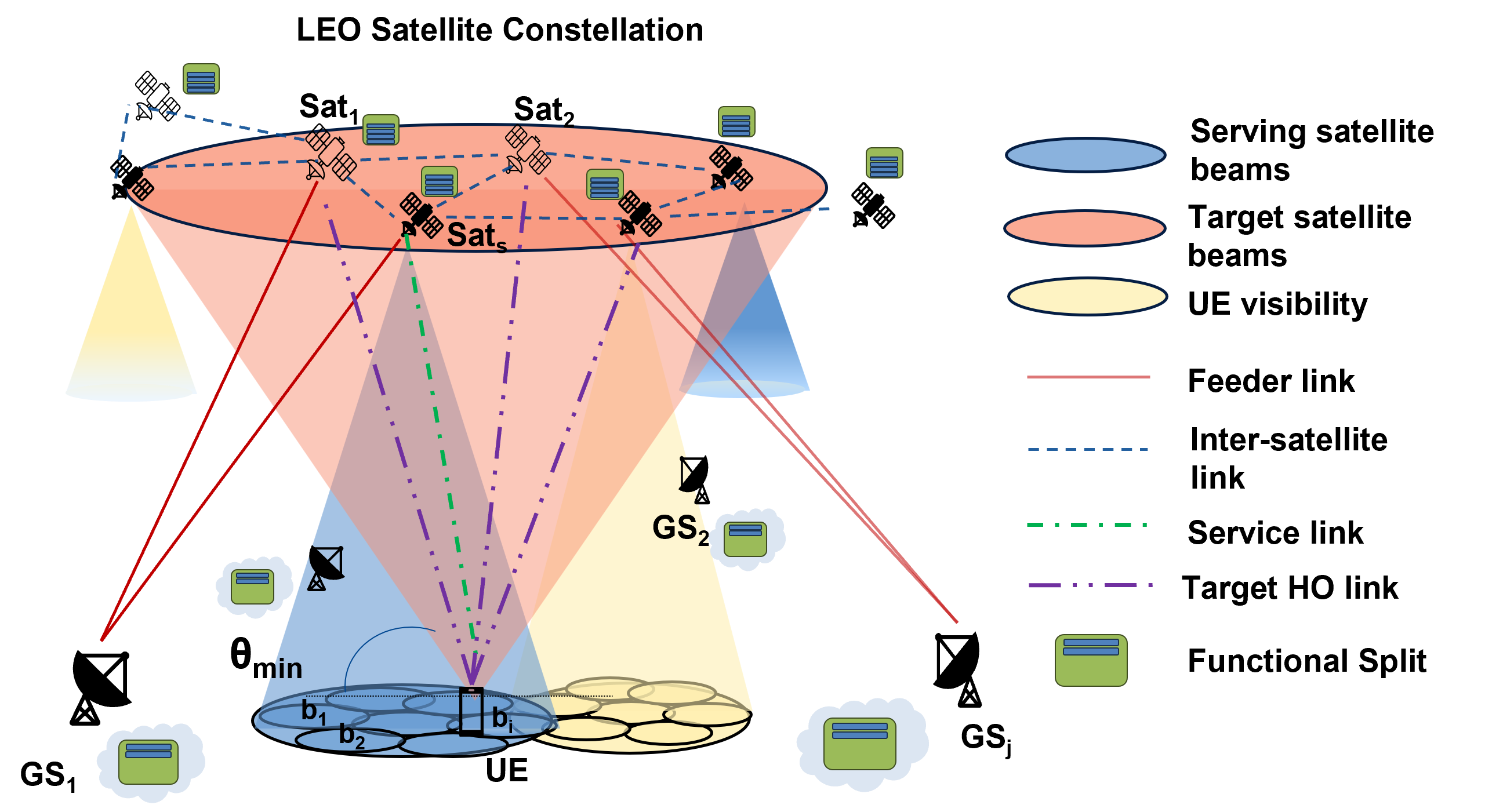}
\caption{System model.}
\label{fig:system model}
\vspace{-6mm}
\end{figure}

As shown in Figure~\ref{fig:system model}, we consider a ground-based UE served by LEO satellites with O-RAN FSs. Each satellite is equipped with highly directional circular aperture antennas~\cite{tr38811} that serve ground UEs using Earth-moving beams (non-steerable beams). The visibility of the UE and GS is determined by their respective elevation angles. The elevation angle is defined as the angle between the horizontal plane at the UE or GS and the line connecting it to the satellite. A UE is considered visible to the satellite if its elevation angle exceeds a predefined minimum threshold, $\theta_{min}$, ensuring an unobstructed line-of-sight (LoS) as shown in Figure~\ref{fig:system model}. Similarly, GS visibility is constrained by a minimum GS elevation angle, $\beta_{min}$. %Here, $\theta_{min}$ is always higher than $\beta_{min}$ due to the limited transmission power of the UE at lower elevation angles~\cite{3GPP_NTN}. %In contrast, GSs are not power-constrained, allowing them to operate with a lower minimum elevation angle. Consequently, UE is positioned such that the UE visibility region is entirely contained within the GS visibility region as shown in Figure~\ref{fig:system model}.
Within this overlapped visibility region, we analyze CHO performance for satellites as they move in and out over time. Let the LEO satellite index be denoted as \( s \in \{ 1,2,\ldots,\text{S}\} \), beam indices be denoted as \( i \in \{ 1,2,\ldots,\text{I}\} \) and the GS indices be denoted as \( j \in \{ 1,2,\ldots,\text{J}\} \). Time slots within the service period are indicated by \( t \in \{ 1, 2, \ldots, \text{T} \} \) and time slot duration is defined as $\Delta t$. The service time of the UE is defined as $\text{T}\Delta t$. The GS association with satellite $s$ through feeder link (FL) at time $t$ is indicated by $g_s(t)$, where $g_s(t)$ can take values from set \{0,1,2,\ldots,\text{J}\}. Here, $0$ indicates no association of satellite $s$ with any GS. Additionally, at time $t$ the elevation angle between UE and satellite $s$ and between GS $j$ and satellite $s$ is denoted by $\theta_{s}(t)$ and $\beta_{s,j}(t)$ respectively. Furthermore, during each time slot, the UE has access to a set of candidate satellites that meet specific visibility conditions. The candidate satellite set $\mathcal{S}$($t$) at time $t$ is defined as
%\begin{align}
%   & \theta_{s}(t) \geq \theta_{min}, \quad \forall t,\forall s, \label{prob: cons3_updated} \\
%   & \beta_{s,j}(t) \geq \beta_{min}, \quad \forall t, \forall s, \forall j. \label{prob: cons4_updated} \ \
%\end{align}
\begin{equation}
    \mathcal{S}(t) = \{ s : \theta_{s}(t) \geq \theta_{min}, \beta_{s,j}(t) \geq \beta_{min}, \forall s, \forall j\}.
    \label{eq:candidate set}
\end{equation}

%The UE association with satellite $s$ via the service link (SL) at time $t$ is represented by \( y(t) = s \) and beam association is represented as \( z(t) = i \). An intra-satellite HO occurs when \( y(t) = y(t-1) \) and  \( z(t) \neq z(t-1) \), indicating that the UE remains connected to the same satellite but the serving beam is changed. Conversely, an inter-satellite HO is characterized by \( y(t) \neq y(t-1) \), signifying a switch in the serving satellite. 

\subsection{Path Loss Model}
The overall pathloss (in dB) is defined as \cite{tr38811}:

\begin{equation}
    PL = PL_g + PL_b + PL_s + PL_e,
    \label{eq: total PL}
\end{equation}
where \( PL_g \) is the attenuation due to atmospheric gases, \( PL_b \) is the basic path loss, \( PL_s \) is due to ionospheric scintillation, and \( PL_e \) represents building penetration loss. For this paper, we assume that the UE is located in an outdoor setting, allowing us to neglect \( PL_e \). The basic path loss, \( PL_b \), is defined as:
\begin{equation}
    PL_b = FSPL + SF + CL,
    \label{eq:basic PL}
\end{equation}
where \( FSPL \) represents the free space path loss, \( SF \) denotes the shadow fading loss, and \( CL \) is the clutter loss. Since the UE is assumed to be outdoors under the line of sight (LoS) conditions, \( CL=0 \). Furthermore, as the environment is considered free of significant obstructions, $SF$ is also neglected. Hence the basic pathloss is mainly due to \( FSPL \), which can be expressed as:
\begin{equation}
    FSPL(d_s(t)) = 32.45 + 20 \log_{10} f_c + 20 \log_{10} d_s(t),
    \label{eq:fspl PL}
\end{equation}
where \( f_c \) represents the carrier center frequency in GHz, and \( d_s(t) \) is the distance between the UE and satellite \( s \) at time slot $t$ in meters.

To simplify our study, we assume clear sky and LoS conditions, neglecting all other losses. Additionally, atmospheric losses are considered negligible and are therefore excluded from the analysis.

Subsequently, the RSRP from beam \( i \) of satellite \( s \) can be calculated %, while neglecting some PLs following our assumptions, 
as shown below \cite{3GPP_NTN}:
\begin{equation}
    P_{s,i}(t) = EIRP(\alpha_{s,i}(t)) + G_r - PL,
    \label{eq:Rxpower}
\end{equation}
where \( EIRP(\alpha_{s,i}) \) is the effective isotropic radiated power (EIRP) from beam \( i \) of satellite \( s \) at time slot $t$ in dBW, \( \alpha_{s,i}(t) \) is the angle between the $i^{\text{th}}$ beam center of satellite $s$ and the UE at time slot $t$, and \( G_r \) denotes the receiver’s antenna gain in dBi.
The EIRP can be computed as:
\begin{equation}
    EIRP(\alpha_{s,i}(t)) = P_s + G_T(\alpha_{s,i}(t)),
    \label{eq:EIRP}
\end{equation}
where \( P_s \) is the transmission power of satellite $s$ in dBW, and \( G_T(\alpha_{s,i}(t)) \) is the transmission antenna gain in dBi, given by:

\begin{equation}
    G_T(\alpha_{s,i}(t)) =
    \begin{cases}
        G_t, \quad \quad \quad \quad \quad \quad \quad \quad \quad  \text{if } \alpha_{s,i}(t) = 0, \\
        G_t + 10\log_{10}\!\left( 
        4 \left| \frac{J_1(\kappa a \sin(\alpha_{s,i}(t)))}%
        {\kappa a \sin(\alpha_{s,i}(t))} \right|^2 
        \right),\\ \quad \quad \quad \quad \quad \quad  \quad \quad \quad \quad \quad \text{otherwise,}%\text{if } \alpha_{s,i}(t) \neq 0 
    \end{cases}
    \label{eq:Antenna gain}
\end{equation}

where $G_t$ is the maximum antenna gain, \( J_1(\cdot) \) is the first-order Bessel function, \( \kappa = \frac{2 \pi f_c}{c} \) is the wave number with \( c \) as the speed of light, \( a \) denotes the radius of the equivalent satellite antenna aperture.

\subsection{Conditional Handover Delay Model} 

We use a legacy RSRP-based HO mechanism where the UE connects to the satellite beam with the highest received power, based on the antenna gain and pathloss as in~\eqref{eq:Rxpower} and~\eqref{eq:Antenna gain}. In this method, both intra- and inter-satellite beams are considered when selecting the optimal beam. Additionally, HO parameters, such as TTT and HOM, are introduced to optimize the HO performance. These parameters play a pivotal role in determining when a UE initiates a HO based on received signal measurements~\cite{irshad}.

\begin{itemize}
    \item TTT ($\delta_t$) is the duration for which a target satellite beam's signal quality (\emph{e.g.,} RSRP) must remain better than the serving satellite beam's signal quality before the HO procedure is triggered. A lower TTT value leads to faster HO decisions, while a higher TTT helps avoid UHOs and PPs.
    
    \item HOM ($\delta_h$) is the offset added to the target satellite beam's signal quality, making it relatively easier or harder to satisfy the condition for triggering an HO. A higher HOM delays HO decisions, reducing the likelihood of UHOs but increasing the risk of service interruptions.

\end{itemize}
\begin{figure}[b]
\centering
\includegraphics[width=0.99\linewidth]{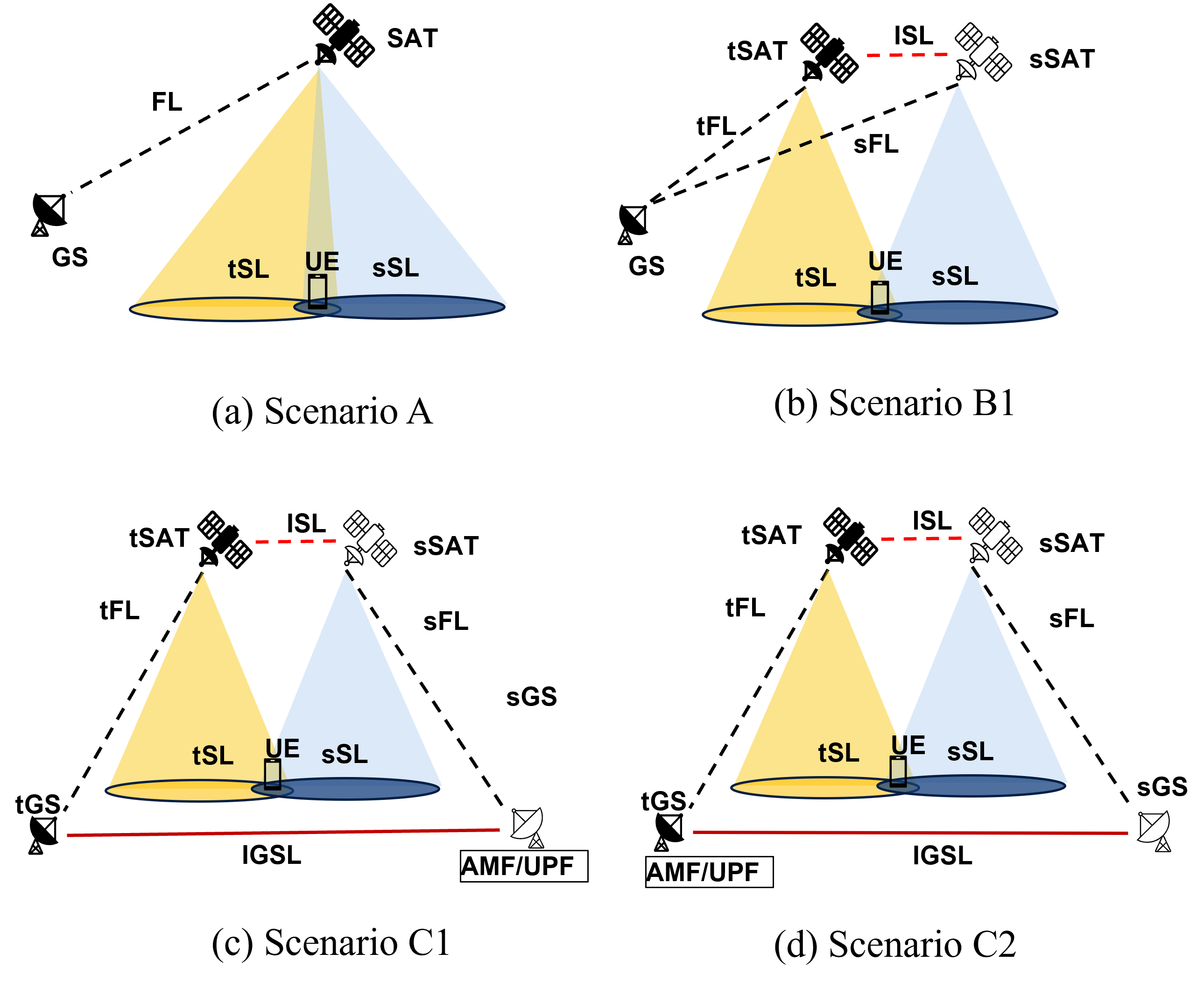}
\caption{Mobility scenarios.}
\label{fig:scenarios}
\end{figure}
According to~\cite{eucnc}, CHO delay is influenced by both the FS deployment and mobility scenarios. We denote the O-RAN FSs by \( f \in \{ 1,2,3\} \), corresponding to split 7.2x, split 2, and gNB onboard the satellite, respectively. The mobility scenarios, depicted in Figure~\ref{fig:scenarios}, are classified as A, B1, C1, and C2~\cite{eucnc}. 

\begin{itemize}
    \item \textbf{Scenario A}: HO occurs between beams of the same satellite, referred to as intra-satellite HO.
    \item \textbf{Scenario B1}: An inter-satellite HO scenario where both the source and target satellites are connected to the same GS via the FL. Here, \( s \) and \( t \) denote the source and target satellites, respectively.
    \item \textbf{Scenario C1}: An inter-satellite HO scenario where the source and target satellites are connected to different GSs, with the access and mobility function (AMF) and user plane function (UPF) hosted at the source GS (sGS).
    \item \textbf{Scenario C2}: Similar to C1, except that the AMF and UPF are hosted at the target GS (tGS).
\end{itemize}

The mobility scenarios are denoted by \( q \in \{ 1, 2, 3, 4 \} \), corresponding to scenarios A, B1, C1, and C2, respectively. The CHO delay refers to the delay experienced by the UE during a HO from one beam to another when the HO condition is triggered. This CHO delay is expressed as a function of \( f \) and  \( q \) as follows:
\begin{equation}
        \tau_{f,q}^{\text{CHO}}(t) =  \ \tau^{\text{sync}} +\textbf{I}\{f=3\}\cdot \textbf{I}\{q\neq 1\} \cdot\tau^{\text{core}}  + \tau_{f,q}^{\text{proc}} + \tau_{f,q}^{\text{prop}}(t),
    \label{eq:delay total dynamic}
\end{equation}
where \( \tau^{\text{sync}} \) is the synchronization delay, \( \tau^{\text{core}} \) is the core network application programming interface (API) call delay, \( \tau_{f,q}^{\text{proc}} \) is the processing delay, \( \tau_{f,q}^{\text{prop}}(t) \) is the propagation delay, and \( \textbf{I}\{\cdot\} \) is an indicator function, equals to $1$ if the condition within parentheses holds and $0$ otherwise.
The processing delay \( \tau_{f,q}^{\text{proc}} \) is modeled as: 
\begin{equation}
    \tau_{f,q}^{\text{proc}} = N_{f,q}^{\text{FL}} \cdot \tau^{\text{ppm}} + N_{f,q}^{\text{SL}} \cdot \tau^{\text{ppm}} + N_{f,q}^{\text{ISL}} \cdot \tau^{\text{ppm}} + N_{f,q}^{\text{IGSL}} \cdot \tau^{\text{ppm}} %) \cdot \tau^{\text{ppm}},
    \label{eq: delay proc dynamic}
\end{equation}
where \( N_{f,q}^{(\cdot)} \) represents the number of message exchanges across various links, including FL, service link (SL), inter-satellite link (ISL), and inter-ground station link (IGSL) for FS $f$ and scenario $q$, while \( \tau^{\text{ppm}} \) denotes the processing delay per message.
The propagation delay \( \tau_{f,q}^{\text{prop}} \) is given by:

\begin{equation}
    \begin{split}
        \tau_{f,q}^{\text{prop}}(t) = & \ N_{f,q}^{\text{FL}} \cdot \tau^{\text{FL}}(t) + N_{f,q}^{\text{SL}} \cdot \tau^{\text{SL}}(t) \\
        & + N_{f,q}^{\text{ISL}} \cdot \tau^{\text{ISL}}(t) + N_{f,q}^{\text{IGSL}} \cdot \tau^{\text{IGSL}}(t),
    \end{split}
    \label{eq:delay prop dynamic}
\end{equation}
where $\tau^{(\cdot)}(t)$ represents propagation delay in different links such as FL, SL, ISL, and IGSL at time slot $t$.

\subsection{Mobility Key Performance Indicators}

\subsubsection{Radio Link Failure (RLF)}

%An RLF occurs at a time slot \( t \) if the RSRP of the serving satellite beam ($P_{s,i}(t)$) falls below the receiver sensitivity threshold ($P_{thr}$) \textit{i.e.,} $P_{s,i}(t)<P_{thr}$. The RLF rate is defined as the number of total RLF events divided by the service time.
An RLF occurs at time slot \( t \) if the serving beam’s RSRP \( P_{s,i}(t) \) falls below the receiver threshold \( P_{thr} \), i.e., \( P_{s,i}(t) < P_{thr} \). The RLF rate is the total number of RLF events divided by the service time.

\subsubsection{Intra-Satellite Handover Rate}

The intra-satellite HO rate is calculated as the ratio of total intra-satellite HOs to the service time ($\text{T}\cdot\Delta t$). It includes occurrences of UHOs and PP events. A UHO occurs when the UE remains connected to the serving beam for less than a predefined duration. A PP event is identified when the UE transitions from beam 1 to beam 2, only to revert to beam 1 after a short interval.

\subsubsection{Inter-Satellite Handover Rate}

The inter-satellite HO rate is defined as the ratio of total inter-satellite HOs to the service time ($\text{T}\cdot\Delta t$). Similar to intra-satellite HOs, this metric encompasses UHO and PP events except that these events happen between satellites.

\subsubsection{Total CHO delay}

The total CHO delay is the cumulative delay of procedures when intra- or inter-satellite HOs are triggered and is given as 
\begin{equation}
    \tau^{tot} =  \sum_{t=1}^\text{T} u(t)\tau_{f,q}^{CHO}(t),
    \label{eq:tot cho}
\end{equation}
where $u(t)=\textbf{I}\big\{P_{s',i'}(t) > P_{s,i}(t) + \delta_h\big\}\cdot \textbf{I}\big\{z<0 \big\}$ defines the conditions for triggering a HO: the RSRP received from a candidate satellite \( s'\in \mathcal{S}(t) \) and beam \( i' \) must consistently exceed the current satellite beam’s RSRP by the HOM \( \delta_h \) for a duration of \( \delta_t \). This is controlled by the timer variable \( z \), as detailed in Algorithm~\ref{alg:exhaustive}. The normalized total CHO delay is defined as the ratio of total CHO delay ($\tau^{tot}$) to the service time ($\text{T}\cdot\Delta t$).
\subsubsection{Effective service time}
The effective service time for a UE can be expressed as 
\begin{equation}
    \xi =  \sum_{t=1}^T r(t)\Delta t - \tau^{tot}=\sum_{t=1}^T( r(t)\Delta t - u(t)\tau_{f,q}^{CHO}(t)),
    \label{eq:effective service time}
\end{equation}
where the first term represents duration without RLF and $r(t)=\textbf{I}\big\{ P_{s,i}(t) \geq P_{thr} \big\}$ is a RLF indicator ensures that the UE is served by a satellite beam only if the received signal power \( P_{s,i}(t) \) exceeds the predefined threshold \( P_{thr} \); otherwise, a RLF occurs, indicated by \( r(t) = 0 \). Availability (normalized effective service time) is defined as the ratio of effective service time ($\xi$) to the service time ($\text{T}\cdot\Delta t$).

\begin{algorithm}[b!]
\caption{Exhaustive Search for Optimal TTT and HOM}
\label{alg:exhaustive}
\begin{algorithmic}[1]
\STATE \textbf{Input:} $S$, $I$, $J$, $\Delta t$, $T$, $P_{thr}$, $f$, TTT range, HOM range, $g_s(t)$ and $\mathcal{S}(t), \quad \forall t \in \{1,\ldots T \} $, 
\STATE \textbf{Output:}  \( \delta_t^{\text{opt}}, \delta_h^{\text{opt}} \) and $\xi^{\text{max}}$
\STATE Initialize \( \xi^{\text{max}} \leftarrow 0 \), \( \delta_t^{\text{opt}} \leftarrow 0 \), \( \delta_h^{\text{opt}} \leftarrow 0 \)
%\STATE Define the range of \( \delta_t \) (TTT) and \( \delta_h \) (HOM)
\FOR{each \( \delta_t \) in TTT range}
    \FOR{each \( \delta_h \) in HOM range}
        \STATE Initialize \( \xi \leftarrow 0 \), 
        \STATE Initialize HO trigger timer: \( z \leftarrow \delta_t \)
        \FOR{\( t = \{1,\ldots T \} \)}
            \STATE Initialize indicators: \( u(t)\leftarrow 0 \), \( r(t)\leftarrow 1 \)
            %\STATE Compute visible satellites and their beam gains
            \STATE Identify the best satellite $s'\in\mathcal{S}(t)$ 
            \STATE and beam $i'$ based on RSRP
            \IF{$P_{s',i'}(t)>P_{s,i}(t)+\delta_h$}
                \STATE Update \( z \leftarrow z -\Delta t\)
                \IF{$z < 0$}
                    \STATE Update \( u(t)\leftarrow 1 \)
                    \STATE Update \(z \leftarrow \delta_t \)
                \ENDIF
            \ENDIF
            \IF{$P_{s,i}(t)<P_{thr}$}
                \STATE Update \( r(t)\leftarrow 0 \)
            \ENDIF
            \STATE Update \( \xi \leftarrow \xi + r(t)\Delta t - u(t)\tau_{f,q}^{CHO}(t) \) 
        \ENDFOR
        %\STATE Normalize performance metrics based on total time slots
        \IF{\( \xi > \xi^{\text{max}} \)}
            \STATE Update \( \xi^{\text{max}} \leftarrow \xi \), \( \delta_t^{\text{opt}} \leftarrow \delta_t \), \( \delta_h^{\text{opt}} \leftarrow \delta_h \)
        \ENDIF
    \ENDFOR
\ENDFOR
\RETURN Optimal parameters: \( \delta_t^{\text{opt}}, \delta_h^{\text{opt}} \)
\end{algorithmic}
\end{algorithm}

\begin{figure*}
     \centering
     \begin{subfigure}[b]{0.32\textwidth}
         \centering
         \includegraphics[width=\textwidth,trim=2.2mm 1.3mm 9.5mm 6mm, clip=true]{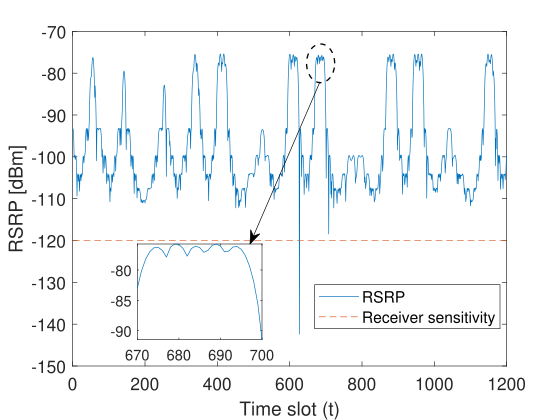}
         \caption{RSRP $\delta_t$=$0$ s, $\delta_h$=$0$ dB}
         \label{fig:RSRP 19 00}
     \end{subfigure}
     \hfill
     \begin{subfigure}[b]{0.32\textwidth}
         \centering
         \includegraphics[width=\textwidth,trim=2.2mm 1.3mm 9.5mm 6mm, clip=true]{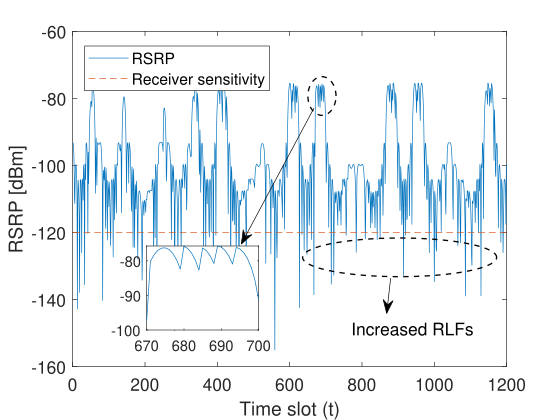}
         \caption{RSRP $\delta_t$=$5$ s, $\delta_h$=$0$ dB}
         \label{fig:RSRP 19 30}
     \end{subfigure}
     \hfill
     \begin{subfigure}[b]{0.32\textwidth}
         \centering
         \includegraphics[width=\textwidth,trim=2.2mm 1.3mm 9.5mm 6mm, clip=true]{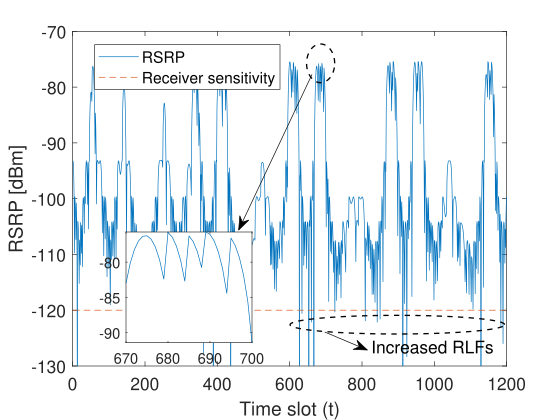}
         \caption{RSRP $\delta_t$=$0$ s, $\delta_h$=$3$ dB}
         \label{fig:RSRP 19 05}
     \end{subfigure}
     \medskip
     \begin{subfigure}[b]{0.32\textwidth}
         \centering
         \includegraphics[width=\textwidth,trim=2.2mm 1.3mm 9.5mm 6mm, clip=true]{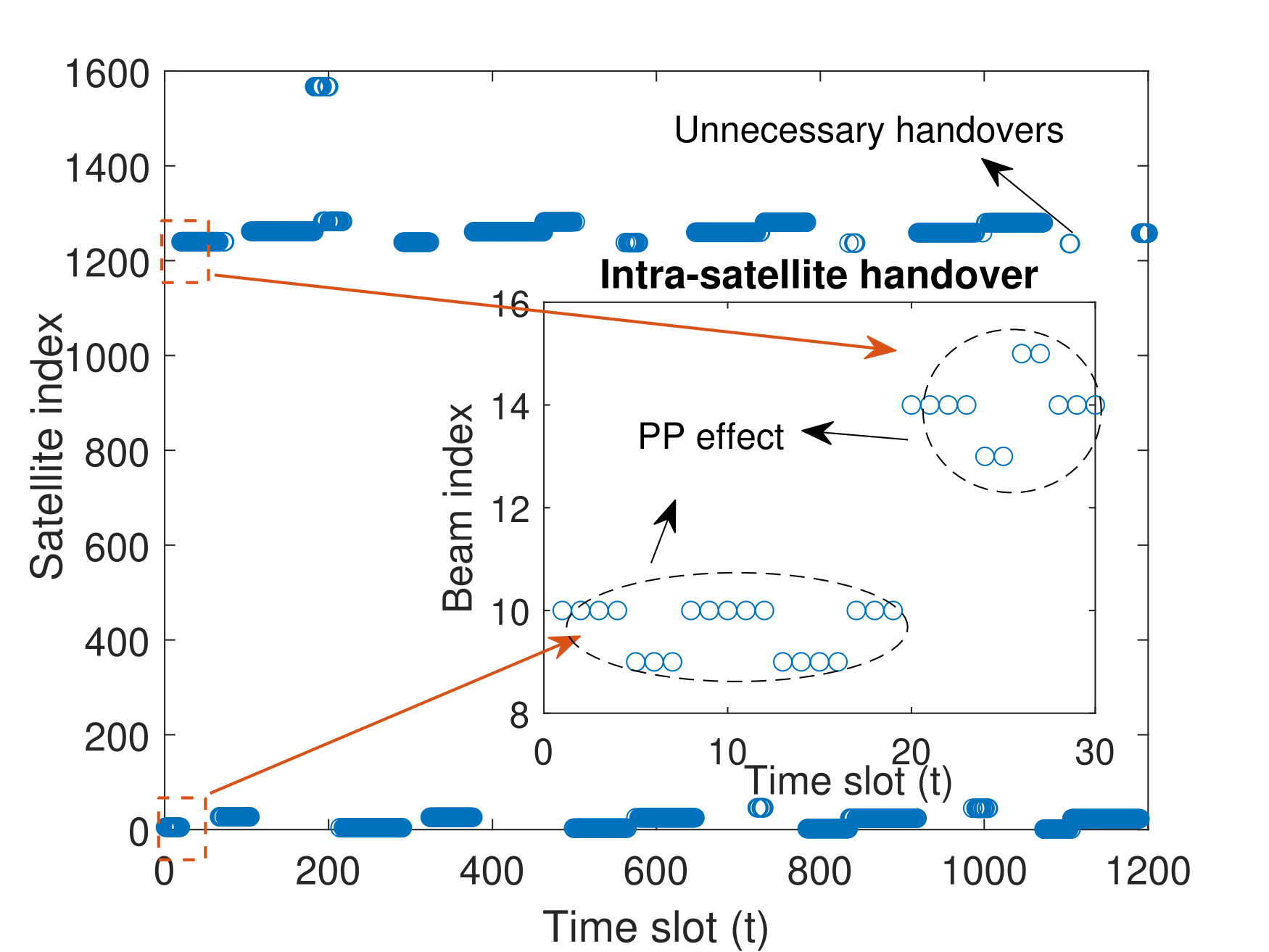}
         \caption{Handover $\delta_t$=$0$ s, $\delta_h$=$0$ dB}
         \label{fig:HO 19 00}
     \end{subfigure}
     \hfill
     \begin{subfigure}[b]{0.32\textwidth}
         \centering
         \includegraphics[width=\textwidth,trim=2.2mm 1.3mm 9.5mm 6mm, clip=true]{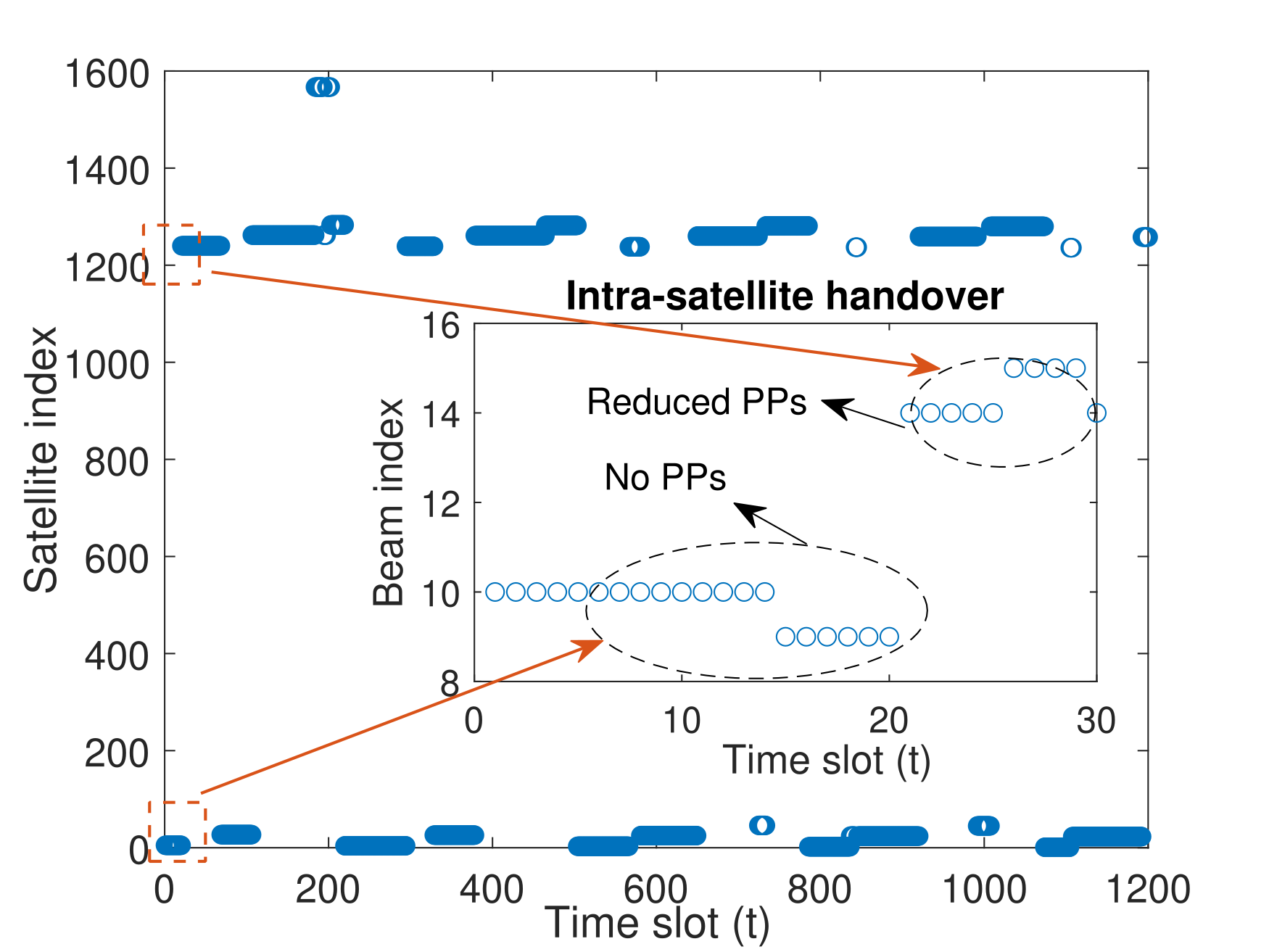}
         \caption{Handover $\delta_t$=$5$ s, $\delta_h$=$0$ dB}
         \label{fig:HO 19 30}
     \end{subfigure}
     \hfill
     \begin{subfigure}[b]{0.32\textwidth}
         \centering
         \includegraphics[width=\textwidth,trim=2.2mm 1.3mm 9.5mm 6mm, clip=true]{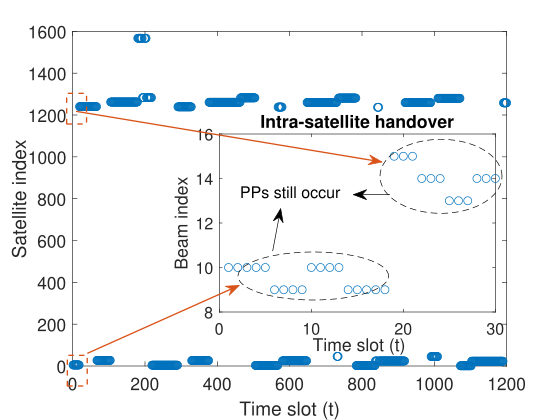}
         \caption{Handover $\delta_t$=$0$ s, $\delta_h$=$3$ dB}
         \label{fig:HO 19 05}
     \end{subfigure}
        \caption{LEO satellites with 19 beam configuration.}
        \vspace{-6mm}
        \label{fig:19 beam configuration}
\end{figure*}
%\vspace{-4mm}
\section{Problem Formulation}
\label{sec:prob form}
The objective is to maximize the effective service time \( \xi \) of the UE by optimizing the HO parameters: \( \delta_t \) and \( \delta_h \) for varying O-RAN FSs and beam configurations. The optimization problem is defined as follows:
\begin{align}
   &\max_{\delta_t, \delta_h}
   \begin{aligned}[t]
      &  \quad   \xi = \sum_{t=1}^T \bigg( r(t)\Delta t - u(t)\tau_{f,q}^{CHO}(t) \bigg), \label{prob: obj_updated}
   \end{aligned}  \\
   \textit{s.t.} \ \  
   & r(t) = \textbf{I}\big\{ P_{s,i}(t) \geq P_{thr} \big\}, \quad \forall t\in \{1,\ldots T \}, \label{prob: cons1_updated} \\
   & u(t) = \textbf{I}\big\{P_{s',i'}(t) > P_{s,i}(t) + \delta_h\big\}\cdot \textbf{I}\big\{z<0 \big\}, \quad \notag\\
   & \quad\quad\quad\quad\quad\quad\quad\quad \forall t\in \{1,\ldots T \},\forall s'\in \mathcal{S}(t),\label{prob: cons2_updated}
   %& \theta_{s}(t) \geq \theta_{min}, \quad \forall t,\forall s, \label{prob: cons3_updated} \\
   %& \beta_{s,j}(t) \geq \beta_{min}, \quad \forall t, \forall s, \forall j, \label{prob: cons4_updated} \ \
\end{align}
where \eqref{prob: cons1_updated} ensures that the UE is served by a satellite beam only if the received signal power exceeds the predefined threshold; otherwise, a RLF occurs, indicated by $r(t) = 0$. \eqref{prob: cons2_updated} defines the conditions for triggering a HO, where $u(t)=1$ indicates the HO occurred at time slot $t$ and $u(t)=0$ otherwise where $\delta_h$ is the HOM variable and $z$ is the timer for TTT variable $\delta_t$.
%imposes the minimum elevation angle that a UE can connect to a satellite $s$ and \eqref{prob: cons2_updated} imposes the minimum elevation angle \( \beta_{min} \) for the GS $j$ to connect to satellite $s$ ensuring robust ground-to-satellite links.

Maximizing the objective function in ~\eqref{prob: obj_updated} is indirectly performing the joint minimization of the RLF duration and total CHO delay. Both of these metrics are influenced by the indicators $u(t)$ and $r(t)$, as defined in~\eqref{eq:tot cho} and~\eqref{eq:effective service time}, respectively. These indicators, in turn, depend on the HO parameters $\delta_h$ and $\delta_t$ as elaborated in Algorithm~\ref{alg:exhaustive}. To solve this simple optimization problem, we use an exhaustive search algorithm by varying the optimization variables $\delta_t$ and $\delta_h$. We define discrete value ranges and granularities of decision variables $\delta_t$ and $\delta_h$ based on practical constraints and deployment-specific requirements. These ranges should be chosen carefully to balance computational efficiency and coverage of potential optimal values. Although an exhaustive search guarantees finding the global optimum, it can be computationally intensive. Thus, we may choose coarser granularity in parameter values. The computational complexity of the exhaustive search algorithm is $\mathcal{O}(M\times N)$, where $M$ and $N$ correspond to the number of grid points for the TTT and HOM ranges. The exhaustive search algorithm is a straightforward but effective approach, particularly in cases with limited parameters or where computational resources allow for a comprehensive search. For our NTN scenario, it provides insights into how varying TTT and HOM values impact HOs and effective service time.%, guiding parameter tuning in real-world deployments.

\begin{table}[h]
\centering
\caption{Simulation parameters.}
\label{tab:parameters}
\begin{tabular}{|c|c|}
\hline
\textbf{Parameter} & \textbf{Values} \\ \hline
$\tau_{\text{sync}}$ & $20$ ms\\ \hline
 $\tau_{\text{core}}$ &  $50$ ms\\ \hline
$\tau_{\text{ppm}}$ & $1$ ms\\ \hline
UE location & ${[}45.78^\circ \text{N}, 1.75^\circ \text{E}{]}$ \\ \hline
$\text{T}$ & $1200$\\ \hline
$\Delta t$ & $1$ s \\ \hline
%Service time ($T\Delta t$) & $1200$ s\\ \hline
$a$ & $0.1$ m \\ \hline
Satellite beam diameter & $50$ km \\ \hline
$P_s$ & $23$dBW \\ \hline
$G_t$ & $30.5$ dBi \\ \hline
$G_r$ & $39.7$ dBi \\ \hline
$P_{thr}$ & $-120$ dBm \\ \hline
$f_c$ & $20$ GHz (Ka-Band) \\ \hline
$\delta_h$ range & [$0$, $15$] dB \\ \hline
$\delta_h$ granularity & $5$ dB \\ \hline
$\delta_t$ range & [$0$, $30$] s \\ \hline
$\delta_t$ granularity & $3$ s  \\ \hline
\end{tabular}
\end{table}

\section{Results and Analysis}

\label{sec:results}
The system-level simulation setup detailed in Section~\ref{sec:system model} is implemented following recommendations from \cite{3GPP_NTN,tr38811}. The main parameters for our setup are outlined in Table~\ref{tab:parameters}. For constellation dynamics, the Starlink LEO configuration is used, comprising $1584$ satellites in $72$ orbital planes, with $22$ satellites per plane at an altitude of $550$ km and inclination of $53^\circ$~\cite{starlinkstup}. Each satellite has beam configurations of $\text{I}=\{19,127\}$, distributed on the ground in $3$ and $6$ concentric tiers as outlined in~\cite{beam1,beam2,tr38811}. Each beam covers a ground area with a diameter of $50$ km and the satellite specifications are based on the assumptions in~\cite{3GPP_NTN}. Starlink GS locations are incorporated~\cite{GSlocation}, with the number of GSs $\text{J}=4$. The GS positions are: ${[}38.33458^\circ \text{N}, 0.4909^\circ \text{E};$ $ 50.9871^\circ \text{N}, 2.1255^\circ \text{E}; $ $45.3207^\circ \text{N}, 9.1886^\circ \text{E}; $ $ 50.3353^\circ \text{N}, 8.5320^\circ \text{E}{]}$. The UE location is positioned to encompass the selected GSs, with coordinates listed in Table~\ref{tab:parameters}. The satellite associates with a GS upon entering the visibility of the UE, based on the criterion of the highest GS elevation angle. Each GS can simultaneously serve up to two satellites through FLs. If the satellite with the highest elevation angle finds the best GS occupied, it associates with the GS offering the next highest elevation angle. The simulation time is set to $20$ minutes, during which the high density of the LEO constellation ensures a significant number of candidate satellites are visible to the UE at any given moment. Since each satellite remains visible to the UE for around a minute, the $20$-minute duration captures enough satellite interactions to produce reliable results.

The simulation mobility scenarios include A, B1, C1, and C2. To enable C1 and C2 scenarios, the AMF and UPF are hosted at specific GS locations. In this study, GS $1$ and GS $4$ are selected to host the AMF/UPF based on IGSL distance. RSRP-based HO mechanism is implemented for HO selection.

\begin{figure*}
     \centering
     \begin{subfigure}[b]{0.32\textwidth}
         \centering
         \includegraphics[width=\textwidth,trim=2.2mm 1.3mm 9.5mm 6mm, clip=true]{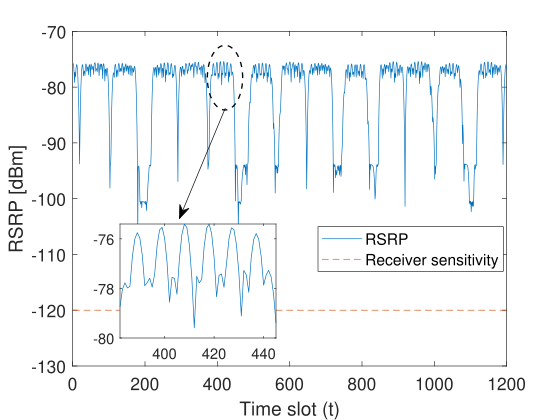}
         \caption{RSRP $\delta_t$=$0$ s, $\delta_h$=$0$ dB}
         \label{fig:RSRP 127 00}
     \end{subfigure}
     \hfill
     \begin{subfigure}[b]{0.32\textwidth}
         \centering
         \includegraphics[width=\textwidth,trim=2.2mm 1.3mm 9.5mm 6mm, clip=true]{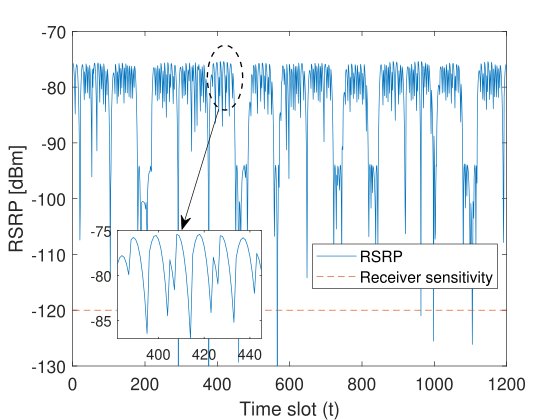}
         \caption{RSRP $\delta_t$=$5$ s, $\delta_h$=$0$ dB}
         \label{fig:RSRP 127 30}
     \end{subfigure}
     \hfill
     \begin{subfigure}[b]{0.32\textwidth}
         \centering
         \includegraphics[width=\textwidth,trim=2.2mm 1.3mm 9.5mm 6mm, clip=true]{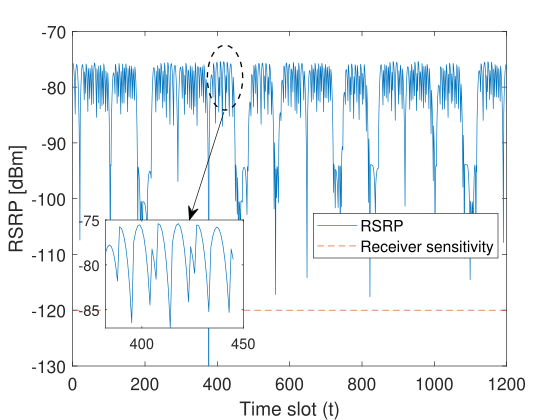}
         \caption{RSRP $\delta_t$=$0$ s, $\delta_h$=$3$ dB}
         \label{fig:RSRP 127 05}
     \end{subfigure}
     \medskip
     \begin{subfigure}[b]{0.32\textwidth}
         \centering
         \includegraphics[width=\textwidth,trim=2.2mm 1.3mm 9.5mm 6mm, clip=true]{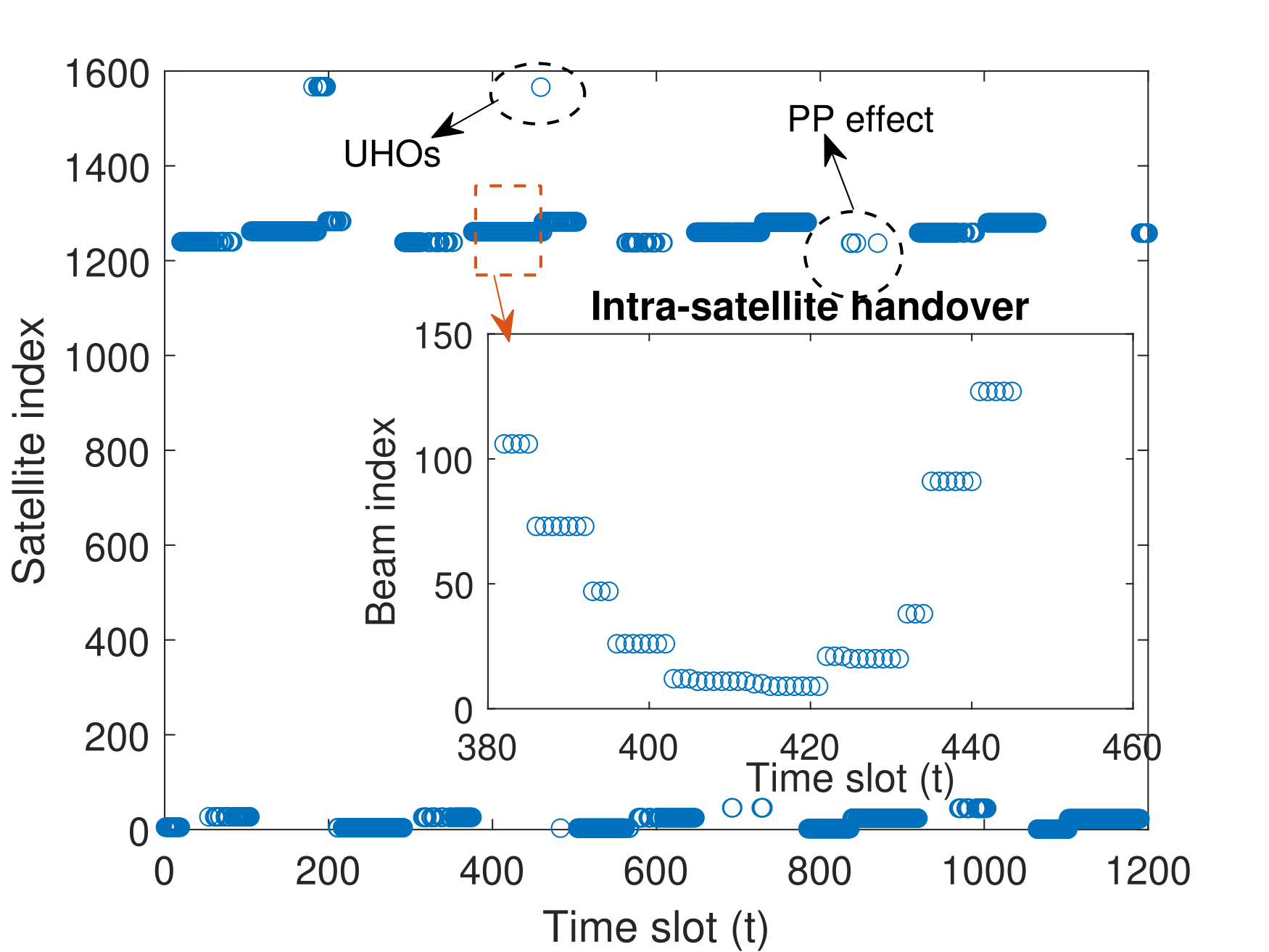}
         \caption{Handover $\delta_t$=$0$ s, $\delta_h$=$0$ dB}
         \label{fig:HO 127 00}
     \end{subfigure}
     \hfill
     \begin{subfigure}[b]{0.32\textwidth}
         \centering
         \includegraphics[width=\textwidth,trim=2.2mm 1mm 9.5mm 6mm, clip=true]{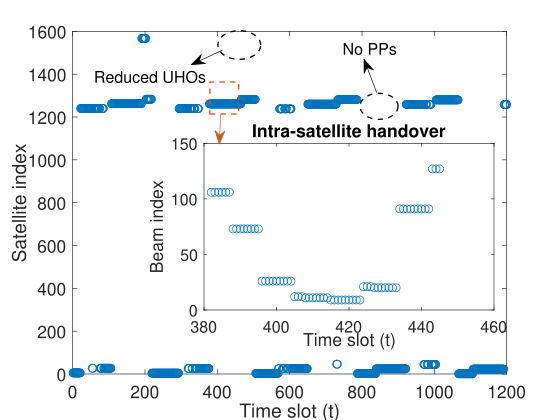}
         \caption{Handover $\delta_t$=$5$ s, $\delta_h$=$0$ dB}
         \label{fig:HO 127 30}
     \end{subfigure}
     \hfill
     \begin{subfigure}[b]{0.32\textwidth}
         \centering
         \includegraphics[width=\textwidth,trim=2.2mm 1.3mm 9.5mm 6mm, clip=true]{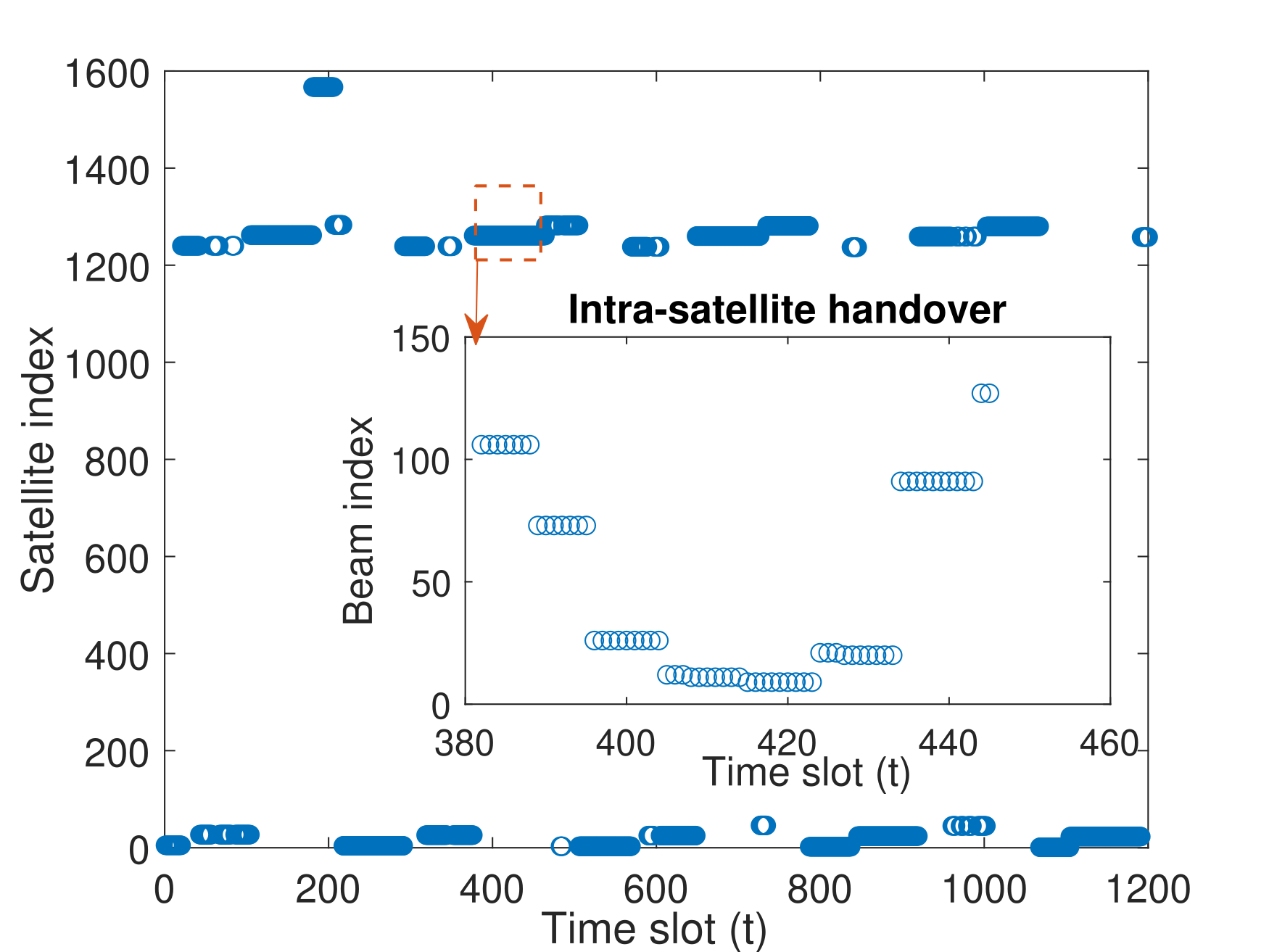}
         \caption{Handover $\delta_t$=$0$ s, $\delta_h$=$3$ dB}
         \label{fig:HO 127 05}
     \end{subfigure}
        \vspace{-3mm}
        \caption{LEO satellites with 127 beam configuration.}
        \label{fig:127 beam configuration}
        \vspace{-4mm}
\end{figure*}
\begin{figure*}
     \centering
     \begin{subfigure}[b]{0.32\textwidth}
         \centering
         \includegraphics[width=\textwidth,trim=2.2mm 1.3mm 5mm 6mm, clip=true]{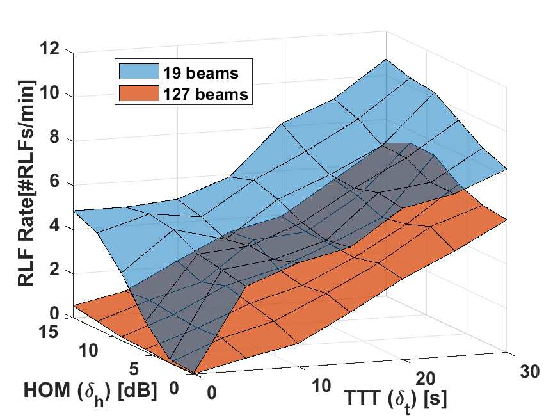}
         \caption{RLF rate vs. HOM and TTT. }
         \label{fig:RLF}
     \end{subfigure}
     \hfill
     \begin{subfigure}[b]{0.32\textwidth}
         \centering
         \includegraphics[width=\textwidth,trim=2.2mm 1.3mm 5mm 6mm, clip=true]{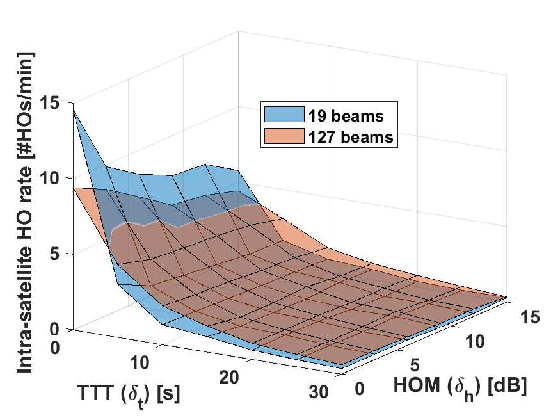}
         \caption{Intra-satellite HO rate vs. HOM and TTT.}
         \label{fig:intra}
     \end{subfigure}
     \hfill
     \begin{subfigure}[b]{0.32\textwidth}
         \centering
         \includegraphics[width=\textwidth,trim=2.2mm 1.3mm 5mm 6mm, clip=true]{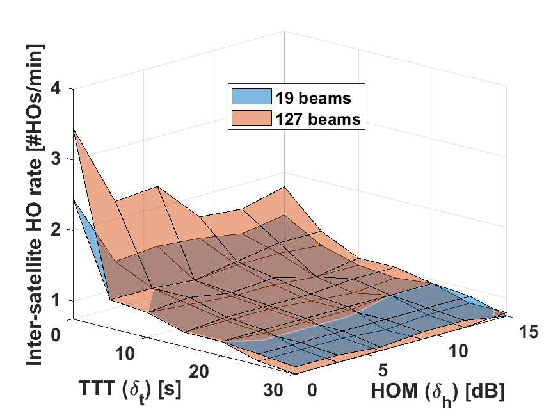}
         \caption{Inter-satellite HO rate vs. HOM and TTT.}
         \label{fig:inter}
     \end{subfigure}
        \vspace{-1mm}
        \caption{HO performance metrics with varying HO parameters TTT and HOM.}
        \label{fig:HO metrics combined}
        \vspace{-4mm}
\end{figure*}
In Figures~\ref{fig:RSRP 19 00}-\ref{fig:RSRP 19 05}, the received RSRP at the UE for a 19-beam configuration is plotted over time influenced by different HO parameters: TTT and HOM. Figures~\ref{fig:HO 19 00}-\ref{fig:HO 19 05} display the corresponding inter-satellite HO patterns, with intra-satellite HOs represented in inset plots. Specifically, Figure~\ref{fig:RSRP 19 00} shows the RSRP received by the UE with $\delta_t = 0$ s and $\delta_h = 0$ dB, resulting in immediate HOs whenever a stronger satellite beam is detected, thereby minimizing RLFs. %Specifically, Figure~\ref{fig:RSRP 19 00} shows the RSRP received by the UE when $\delta_t$ = $0$ s and $\delta_h$ = $0$ dB, which leads to immediate handovers whenever a stronger satellite beam is detected. Consequently, this minimizes RLFs. 
However, with $\delta_t$ = $5$ s, RLFs increase significantly due to the restriction of HO to the optimal beam for a certain duration. Likewise, setting $\delta_h$ = $3$ dB increases RLFs by restricting HOs to a new beam until it offers a significantly stronger signal by a margin than the current one.

To further examine the HO behavior in 19 beam configuration, Figure~\ref{fig:HO 19 00} illustrates inter-satellite HOs across the simulation, with intra-satellite HOs displayed in an inset plot to provide insight into beam-specific HO patterns. For $\delta_t$ = $0$ s and $\delta_h$ = $0$ dB, intra-satellite HOs show a pronounced PP effect, while some inter-satellite HOs result in UHOs. Increasing $\delta_t$, in Figure~\ref{fig:HO 19 30}, reduces the PP effect, and an increased $\delta_h$ similarly reduces UHOs, as in Figure~\ref{fig:HO 19 05}.

An analogous analysis for a 127-beam configuration is shown in Figure~\ref{fig:127 beam configuration}. Figures~\ref{fig:RSRP 127 00}-\ref{fig:RSRP 127 05} reveal that the trend in RLFs with varying $\delta_t$ and $\delta_h$ is consistent with the 19-beam configuration. However, with a 127-beam configuration, the UE is covered by the same satellite for more duration due to the increased number of beams. Regarding HO patterns, the PP effect is less prevalent in intra-satellite HOs for the 127-beam setup, yet it remains observable in inter-satellite HOs, as illustrated in Figure~\ref{fig:HO 127 00}. These UHOs and PP effect diminishes as $\delta_t$ and $\delta_h$ increase, as shown in Figures~\ref{fig:HO 127 30} and \ref{fig:HO 127 05}. Additionally, intra-satellite HOs occur every few seconds, as highlighted in the inset plots of Figures~\ref{fig:HO 127 00}-\ref{fig:HO 127 05}, emphasizing one of the key disadvantages of using earth-moving cells, as compared to earth-fixed cells.

Having established the influence of $\delta_t$ and $\delta_h$ on RLF and HO characteristics, Figure~\ref{fig:RLF} further explores RLF rates for the 19-beam and 127-beam configurations with varying HO parameter values. When $\delta_t$ and $\delta_h$ are low, both configurations experience reduced RLF due to immediate HOs to the optimal satellite beam. However, as these parameters increase, the UE remains connected to the same satellite for longer durations, leading to higher RLF rates, peaking at approximately $11$ RLFs/min for the 19-beam and $7.5$ RLFs/min for the 127-beam configurations. The higher RLF rate in the 19-beam results from its smaller footprint compared to the 127-beam.

In Figure~\ref{fig:intra}, intra-satellite HO rates for both configurations are presented under varying HO parameters. At $\delta_t$ = $0$ s and $\delta_h$ = $0$ dB, the intra-satellite HO rate is high, approximately $15$ HOs/min for the 19-beam and $10$ HOs/min for the 127-beam, reflecting the frequent PP effect in the former, as shown in Figure~\ref{fig:HO 19 00}. When $\delta_t$ exceeds $5$ s, regardless of $\delta_h$, the HO rate in the 127-beam configuration slightly exceeds that of the $19$-beam configuration, as seen in Figure~\ref{fig:intra}. This is because the smaller footprint in 19-beam setup may no longer cover the UE, while in the 127-beam setup, the satellite footprint can still maintain coverage, supporting intra-satellite HO even after $5$ s.

Figure~\ref{fig:inter} illustrates the inter-satellite HO rates for both configurations. At $\delta_t = 0$ s and $\delta_h = 0$ dB, the 127-beam configuration exhibits a higher HO rate (approximately 3.5 HOs/min) compared to the 19-beam configuration (2.5 HOs/min). This is primarily due to the increased occurrence of UHOs and PPs, which are more prevalent in the 127-beam configuration due to the greater multi-coverage of satellites, as shown in Figure~\ref{fig:HO 127 00}. However, as $\delta_t$ and $\delta_h$ increase, the HO rates converge between the two configurations. At $\delta_t$ values above $20$ s, regardless of $\delta_h$, the HO rate in the $19$-beam configuration slightly surpasses that of the $127$-beam configuration, as depicted in Figure~\ref{fig:inter} due to the high RLF rate of 19-beam setup at higher $\delta_t$ shown in Figure~\ref{fig:RLF}.

\begin{figure}[t]
\centering
\includegraphics[width=0.89\linewidth]{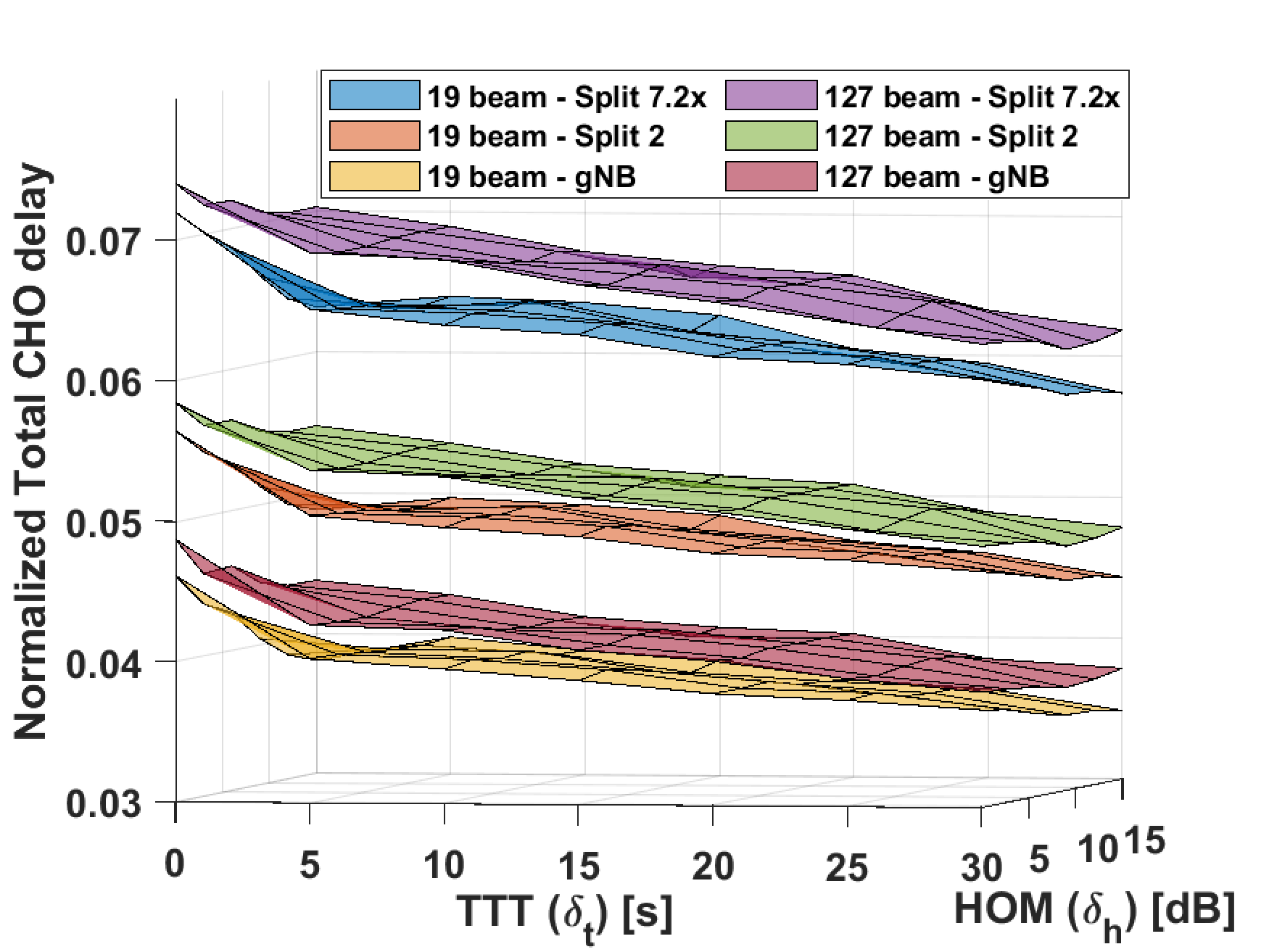}
\caption{Normalized total CHO delay vs. TTT and HOM.}
\label{fig:CHO FS}
\vspace{-4mm}
\end{figure}

\begin{figure}[t]
\centering
\includegraphics[width=0.89\linewidth]{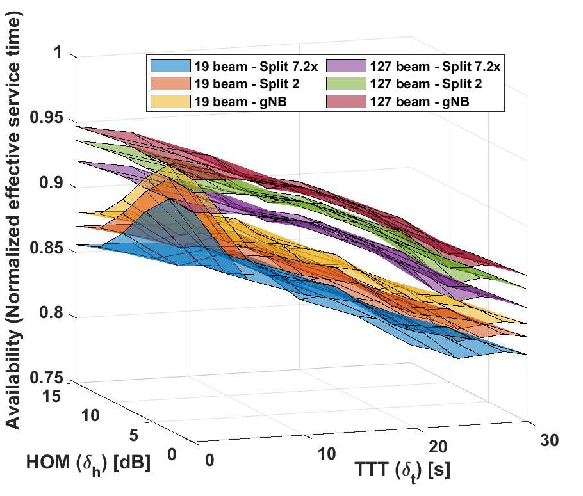}
\caption{Availability vs. TTT and HOM.}
\label{fig:servicetime}
\vspace{-4mm}

\end{figure}

Among  FSs, the gNB onboard the satellite has the minimum total CHO delay ($4.5\%$--$5\%$), as all RAN functions are on satellite, while the split 7.2x and split 2 show higher total CHO delays, around $7\%$--$7.5\%$ and $5.5\%$--$6\%$, respectively, as depicted in Figure~\ref{fig:CHO FS}. % Although the gNB onboard the satellite offers the minimum total CHO delay, it requires high satellite processing capacity, while the split 7.2x, although advantageous for satellite processing requirements, demands substantial fronthaul bandwidth~\cite{eucnc}. 
In Figure~\ref{fig:servicetime}, the results of the exhaustive search algorithm are shown, illustrating the variation of the availability, with respect to changes in TTT \( (\delta_t) \) and HOM \( (\delta_h) \) for different O-RAN FSs and beam configurations. We found that the optimal HO parameters are \( \delta_t^{\text{opt}} = 0 \) s and \( \delta_h^{\text{opt}} = 3 \) dB across all FSs and beam configurations. This moderate increase in HOM to \( \delta_h = 3 \) dB enhances the UE's availability compared to \( \delta_h = 0 \) dB, mainly by reducing the HO rate (by reducing UHO and PP), which decreases the total CHO delay and thereby improves availability. Additionally, Figure~\ref{fig:RLF} shows that the RLF rate is almost zero at \( \delta_h = 3 \) dB, which further contributes to maximizing the availability. The optimal \( \delta_t^{\text{opt}} = 0 \) s indicates that immediate HO decisions yield the highest availability. This finding underscores that due to the short satellite visibility durations and rapidly changing link quality, delaying HO decisions may lead to suboptimal performance. However, this outcome is specific to the simulation parameters and the LEO altitude considered; in scenarios with higher satellite altitudes or slower mobility dynamics, a non-zero TTT might prove beneficial for mitigating UHOs.

%Nevertheless, the existence of the TTT parameter remains essential in broader contexts. For example, in terrestrial networks or NTN scenarios with slower-moving (high altitude) satellites or less dynamic environments, a non-zero TTT can effectively mitigate unnecessary HOs and improve overall network stability. Thus, while \( 0 \) is the optimal TTT for the conditions considered in this work, its presence as a tunable parameter ensures adaptability across diverse network scenarios.

Overall, the gNB onboard the satellite in both the 127-beam and 19-beam configurations achieves the highest availability at near-optimal HO parameters with $95.4\%$. However, the availability for the 19-beam decreases sharply with increasing HO parameters due to a heightened RLF rate, as observed in Figure~\ref{fig:RLF}. A similar trend is seen for split 2 and split 7.2x, with split 2 offering the second-highest availability of $94.3\%$, followed by split 7.2x with $92.8\%$.

\section{Conclusion}

\label{sec:conclusion}

%This paper analyzes the effects of HO parameters, including TTT and HOM, on RLFs, HO rates with varying beam configurations and FS deployments in LEO satellite constellations. The results show that lower TTT and HOM values reduce RLFs by allowing the UE to connect to the best beam quickly, while higher values introduce delays that increase RLF rates, especially in the 19-beam configuration. The 127-beam configuration, with its higher beam density, generally provides better signal stability and a lower RLF rate, though it requires more frequent inter-satellite HOs when TTT and HOM are low.

%For satellite networks that prioritize low RLF rates and enhanced signal stability, the 127-beam configuration, combined with optimized TTT and HOM settings, provides a robust solution, especially when implemented with a gNB onboard the satellite to reduce CHO delays. In contrast, in scenarios where reducing satellite processing demands and CHO delays is essential, a 19-beam configuration with adjusted TTT and HOM settings and an split 7.2x deployment may be more suitable, despite its relatively higher RLF rate.

This study investigated the impact of various O-RAN FSs and HO parameter configurations on the effective service time of UE in a dynamic and realistic scenario, considering the Starlink LEO constellation and GS HO management. Through an exhaustive search over TTT and HOM parameters, we determined that an optimal setting of \( \delta_t = 0 \) s and \( \delta_h = 3 \) dB consistently maximizes the availability across all FSs and beam configurations. This setting improves the balance between minimizing HO rates and reducing RLF occurrences, resulting in reduced total CHO delay and near-zero RLF rates. %Additionally, the gNB onboard the satellite, which places the entire RAN stack on the satellite, results in the lowest HO delays and consequently the highest effective service time. While the gNB configuration increases satellite complexity compared to other O-RAN FSs, it offers significant benefits in effective service time and fronthaul bandwidth, with the trade-off being higher processing demands on satellite.

Among the O-RAN FSs, the gNB onboard the satellite—where the entire RAN stack resides on the satellite—offers the best performance, resulting in the lowest HO delays and the highest effective service time in dynamic mobility scenarios.  This finding contrasts with our previous work~\cite{eucnc}, which analyzed snapshot mobility scenarios and reported higher inter-satellite HO delays for gNB. The discrepancy arises because, in dynamic mobility scenarios, the overall HO delay is influenced by the frequency and type of HOs. The gNB onboard the satellite benefits due to lower intra-satellite HO delays, which occur more frequently, thereby resulting in the lowest overall HO delay. However, this configuration introduces higher satellite complexity and processing demands. In contrast, split 7.2x and split 2, while reducing satellite complexity and processing requirements, face limitations in terms of increased HO delays making them less suitable for NTN scenarios requiring high service continuity. 

Overall, these findings highlight that the appropriate tuning of HO parameters and careful selection of O-RAN FSs can significantly enhance service continuity and reliability in NTN under varying beam configurations.  Future research should focus on developing advanced HO algorithms that incorporate additional HO parameters and adapt to dynamic NTN environments, considering the 5G timer restrictions. %Additionally, exploring machine learning-driven approaches for adaptive parameter tuning could further enhance HO performance and ensure service continuity in complex NTN scenarios.
\vspace{-1mm}
\section*{Acknowledgement}
\vspace{-1mm}

This work was funded by the CELTIC-NEXT Project, 6G for Connected Sky (6G-SKY), with funding received from the Vinnova, Swedish Innovation Agency and was also supported in part by SSF research center, Sustainable Mobile Autonomous and Resilient 6G SatCom (SMART-6GSAT), under registry number CSG23-0001.

\end{document}